\documentclass{jpsj2}

\title{%
Complex microwave conductivity of Na-DNA powders
}

\author{%
H. Kitano\thanks{E-mail address: hkitano@maeda1.c.u-tokyo.ac.jp}, 
K. Ota, A. Maeda
}

\inst{%
Department of Basic Science, University of Tokyo, 
3-8-1, Komaba, Meguro-ku, 153-8902, Japan
}

\recdate{\today}

\abst{%
We report the complex microwave conductivity, $\sigma=\sigma_1-i\sigma_2$, 
of Na-DNA powders, which was measured from 80~K to 300~K by using 
a microwave cavity perturbation technique. 
We found that the magnitude of $\sigma_1$ near room temperature was much larger than the contribution of the surrounding water molecules, and that 
the decrease of $\sigma_1$ with decreasing temperature was sufficiently stronger than that of the conduction of counterions. 
These results clearly suggest that the electrical conduction of Na-DNA is 
intrinsically semiconductive. 
}

\kword{DNA, complex microwave conductivity, cavity perturbation, electrical conduction, powder}

\begin{document}
\maketitle

\section{Introduction}

The electrical conduction of DNA has attracted much recent interest from 
the viewpoints of the fundamental understanding of the charge transfer 
mechanism along a DNA double helix with $\pi-\pi$ orbital stacking. 
\cite{Eley1962}
However, many experimental results have still been controversial, because of several experimental difficulties.\cite{Fink1999,Porath2000,Yoo2001,Zhang2002,TranPRL00,Briman} 
For instance, there are two serious difficulties in the two-probed dc measurements of DNA wires prepared by a nanometer-scale microfabrication technique.\cite{Fink1999,Porath2000,Yoo2001,Zhang2002} 
One is the difficulty in the direct physical contact between the metal electrode and the DNA sample, and the other is the influence of the ionic conduction due to a residual buffer. 
In addition, the dc measurements are easily influenced by the existence of disorder and defects in DNA wires. 

On the other hand, contactless measurements at the microwave frequencies are free from the contact problem, suggesting a new route toward the understanding of the electrical conduction of DNA. 
However, the previous studies at the microwave and quasi-optical frequencies were faced by another difficulty.\cite{TranPRL00,Briman} That is, the dipole relaxation of water molecules surrounding DNA backbone seriously affects the ac conductivity in the frequency region above 100~GHz, depending on the humidity at which DNA samples are measured. \cite{Briman}

In this paper, by using a microwave cavity perturbation technique, we investigated the electrical conduction of DNA powders, which were sealed in a glass tube together with He gas in order to minimize the influence of dipole relaxation of water molecules. 
By removing the contribution of the glass tube to the microwave response very carefully, we succeeded in obtaining the complex microwave conductivity, $\sigma(=\sigma_1-i\sigma_2)$ of DNA powders, showing a semiconductive behavior at least above 100~K. 
This semiconductive behavior was found to be very close to the intrinsic property of the charge transport of DNA molecules rather than the dipole relaxation of residual water molecules. 

\section{Experimental techniques}

\subsection{Sample preparation}

In order to suppress the influence of water molecules included in DNA samples, 
it is necessary to remove a buffer solution completely. 
DNA powders seem to be more suitable for this purpose rather than a thick film of DNA fibers, since such a film still includes non-negligible water molecules even after vacuum drying. 
Thus, we used sodium salt DNA (Na-DNA) powders \cite{Na-DNA}, which were sealed in glass tubes with He gas by the following two methods (Methods I and II). 
Details of the measured samples are listed in Table I. 
In Method I, an atmosphere of the tube was directly pumped out and exchanged into He gas, after putting a small amount of Na-DNA powders in a glass tube (samples A and B). 
The tube was sealed at about 17~mm above the bottom of the tube by using a burner, while the sample part at the bottom was kept in liquid nitrogen, in order to protect the DNA sample against a heating damage. 
An empty glass tube was also prepared by the same method as a reference. 
These sample tubes prepared by Method I were used to determine the qualitative electric properties of DNA powders. 

On the other hand, Method II was used to obtain the electric conductivity of DNA powders more quantitatively. 
Three glass tubes with the same dimensions (1.2~mm outer diameter and 
12~mm height) were prepared (samples C, D, and E). 
Different amounts of Na-DNA powders were put in them inside a glove box, 
of which the atmosphere was exchanged into He gas. 
A very small amount of an epoxy (Stycast 1266) was used to seal the tubes 
at room temperature, in contrast to Method I. 

In Table I, we estimated a packing fraction of DNA powders, which was necessary to obtain the complex conductivity, by comparing the apparent volume packed in the glass tube with the true volume estimated from the specific gravity 1.625~g/cm$^3$ of dry Na-DNA. \cite{Franklin}

\subsection{Microwave cavity perturbation method}
The microwave response was measured by using a cavity perturbation technique 
\cite{MW-review}. 
As shown in Fig.~1, the sample tube was inserted into a copper cylindrical 
cavity resonator (38~mm of diameter and 30.5~mm of height) 
operating at 10.7~GHz in the TE$_{\rm 011}$ mode, in order to measure 
the change of the resonance frequency, $f$, and that of the dissipation, $1/Q$ (Here, $Q$ is a quality factor). 
To remove a large temperature-dependent background of the cavity resonator, 
the temperature of the cavity resonator was maintained at 77~K in a bath of 
liquid nitrogen. 
By using a hot finger of a sapphire plate to support the sample tube, 
the sample was heated up to 300~K, which was thermally isolated from 
the cavity resonator. The value of loaded $Q$  was typically 2.4$\times$10$^4$ at the base temperature. 

As was previously reported,\cite{KitanoPRL2002} the information on the electric conduction of unknown powders can be obtained by comparing $\Delta(1/2Q)$ (or $\Delta f/f$) at the antinode of the microwave electric field ($E_\omega$) with that at the antinode of the microwave magnetic field ($H_\omega$). 
This is based on the fact that $\Delta(1/2Q)$ and $\Delta f/f$ induced by the sample are determined by the electromagnetic properties of the sample. 
In Fig.~2, we plot the calculated $\Delta(1/2Q)$ and $\Delta f/f$ at $E_\omega$ (or $H_\omega$) as a function of the real part of $\sigma$, $\sigma_1$, for several values of the grain size of powders, $a$, and the real part of the complex dielectric constant, $\epsilon_1$, assuming the measured frequency as 10.7~GHz and the sample volume as 1~mm$^3$. 
Although details of $\Delta(1/2Q)$ (or $\Delta f/f$) at $E_\omega$ depend on the value of $\epsilon_1$, and those at $H_\omega$ depend on $a$, we can obtain the following important features from them. That is, independent of $\epsilon_1$ and $a$, $\Delta(1/2Q)$ at $H_\omega$ is much larger than that at $E_\omega$ in a higher conductive region, while $\Delta(1/2Q)$ at $H_\omega$ is much smaller than that at $E_\omega$ in a lower conductive region. 
Furthermore, in the higher conductive region, the magnitude of $\Delta f/f$ at $H_\omega$ is comparable to that at $E_\omega$, while the sign of $\Delta f/f$ at $H_\omega$ (positive) is different from that at $E_\omega$ (negative). 
On the other hand, in the lower conductive region, the magnitude of $\Delta f/f$ at $H_\omega$ is much smaller than that at $E_\omega$, while the signs at both fields are the same (negative), as shown in Fig.~2(b).

It should be noted that these features are never changed, even if one considers the effect of the distribution of $a$ or the ambiguity of $\epsilon_1$. 
Thus, our method can successfully be applied to any powders in which 
a usual dc resistivity measurement is impossible. 

\subsection{Estimation of complex conductivity}

In general, the procedure to obtain the complex conductivity, 
$\sigma(=\sigma_1-i\sigma_2)$, from both $\Delta(1/2Q)$ and $\Delta f/f$ is 
established in the only two limiting regimes \cite{KleinReview}. 
One is the metallic regime (or the so-called ``skin depth regime"), where 
the microwave response is directly related to a complex surface impedance, 
$Z_s(=\sqrt{i\omega\mu/\sigma})$. 
The other is the insulating or semiconducting regime 
(``depolarization regime"), where the microwave response measured 
at $E_\omega$ is rather related to a complex dielectric constant, 
$\epsilon(=\epsilon_0+4\pi i\sigma/\omega)$. 
As will be mentioned below, we found that the electric conduction of Na-DNA is 
insulating or semiconducting, by comparing the microwave response 
measured at $E_\omega$ with that at  $H_\omega$. 
Based on this observation, we estimated the microwave conductivity 
of Na-DNA powders by using the measured data at $E_\omega$. 
If we define an effective complex dielectric constant, $\epsilon_p$, 
as a correction of the powder form, $\epsilon_p$ will be obtained from 
$\Delta(1/2Q)$ and $\Delta f/f$ measured at $E_\omega$, as follows, 

\begin{equation}
\epsilon_p=1-\frac{1}{n}
\frac{\Delta f/f-i\Delta(1/2Q)}{\Delta f/f+\gamma/n-i\Delta(1/2Q)}. 
\label{e1}
\end{equation}

\noindent
Here, $\gamma$ is a geometrical factor proportional to the volume ratio of 
the sample to the cavity (typically, $\sim$5-10$\times$10$^{-5}$). 
$n$ is the so-called depolarization factor, 
which was estimated to be $\sim$0.35-0.4 by approximating the cylindrical 
shape of the sample part in the sealed tube to a prolate spheroid 
\cite{KleinReview}. 
The complex dielectric constant of the bulk sample, $\epsilon$, 
is given by the so-called B\"{o}ttcher formula \cite{Bottcher}, as 

\begin{equation}
\epsilon=\epsilon_1+i\epsilon_2=
\frac{\epsilon_p(2\epsilon_p+3\delta-2)}{(3\delta-1)\epsilon_p+1}, 
\label{e2}
\end{equation}

\noindent
where $\delta$ is the packing fraction of the sample powders. 
The real part of the complex conductivity, $\sigma_1$, is directly given by the imaginary part of $\epsilon$ ($\sigma_1=\omega\epsilon_2/4\pi$). 

The validity of this formula has been confirmed by an earlier X-band microwave measurements for various powder materials with different particle size and packing fraction, showing a good accuracy for materials with low permittivity ($<10$). \cite{Dube1969}
Recently, this technique was also applied to obtain the microwave conductivity of the ammoniated alkali fullerides, and played an important role to show that the Mott-Hubbard transition in intercalated fullerides is driven by a symmetry reduction of the crystal structure. \cite{KitanoPRL2002}

\section{Results and Discussion}

Figures~3(a) and 3(b) show the temperature dependence of $\Delta(1/2Q)$ and $\Delta f/f$ for sample A, respectively, which were obtained by subtracting the measured data for the empty tube from those for sample A. 
As shown in Fig.~3(a), we found that $\Delta(1/2Q)$ at $E_\omega$ was much larger than that at $H_\omega$ in the whole range of measured temperatures. 
These results clearly suggest that the electric conduction of the measured 
powders is not metallic but insulating or semiconducting. 
The same suggestion was also obtained from the results shown in Fig.~3(b), 
because the sign of $\Delta f/f$ at $H_\omega$ was negative 
and its magnitude was much smaller than that of $\Delta f/f$ at $E_\omega$. 

As was already described in the previous section, in the case of insulating 
or semiconducting materials, the complex microwave conductivity can be obtained from $\Delta(1/2Q)$ and $\Delta f/f$ measured at $E_\omega$. 
In order to obtain the magnitude and the temperature dependence of $\sigma$ more quantitatively, it is necessary to remove the contribution of the glass tube 
from the measured data precisely. 
For this purpose, we compared the results of $\Delta(1/2Q)$ and $\Delta f/f$ for all samples prepared by Methods I and II, as shown in Fig.~4. 
It was clear that the absolute value and the temperature dependence of $\Delta(1/2Q)$ and $\Delta f/f$ for the samples prepared by Method I, which were obtained by subtracting the measured data for the empty tube, showed a spurious behavior due to the slight difference between the glass tubes. 
On the other hand, $\Delta(1/2Q)$ and $\Delta f/f$ for the samples prepared by Method II, which were obtained by subtracting the measured data for the other sample with a smaller amount of DNA powders, were found to show a more quantitatively systematic behavior, suggesting that the contribution of the glass tube could be removed more effectively. 

In addition, we found that $\Delta(1/2Q)$ shown in Fig.~4(c) was rapidly decreased with decreasing temperature, strongly suggesting that the electric conductivity of Na-DNA powders was rapidly decreased with decreasing temperature, as was expected for an insulating or semiconducting material. 
Thus, we concluded that the data of the samples prepared by Method II were more quantitatively close to the intrinsic property of Na-DNA powders. 

Both of $\epsilon_1(T)$ and $\sigma_1(T)$ were obtained by analyzing the results of the samples C to E, as shown in Figs.~5(a) and 5(b), respectively. 
Error bars were estimated by considering the ambiguity of the packing fraction, $\delta$ and of the sample volume in the powder form, $V_{\rm DNA}$. 
We found that the magnitude of $\epsilon_1$ was roughly 3$\sim$8 and that its temperature dependence was very weak between 80~K and 300~K, independent of the samples. 
We confirmed that this value showed a good agreement with a low-frequency dielectric constant of a pelletized sample of Na-DNA, which was measured by HP4192A impedance analyzer in the frequency range from 1~kHz to 50~kHz. 
It seemed that the optical dielectric constant of Na-DNA \cite{Wittlin} was also consistent with our result. 
On the other hand, the magnitude of $\sigma_1$ was estimated to be about 
(1$\sim$3)$\times$10$^{-3}$~$\Omega^{-1}$cm$^{-1}$ at room temperature. 
This value was about three orders of magnitude smaller than the previous estimates for $\lambda$-DNA, which were measured at 12~GHz and 100~GHz using a similar cavity perturbation technique. \cite{TranPRL00} 
We believe that they were overestimates because of the too simplified analysis, where neither the depolarization effect nor the porosity of the DNA samples was considered. 
On the contrary, both effects were appropriately considered in our analysis through Eqs.~(\ref{e1}) and (\ref{e2}). 
The good agreement of the estimate of $\epsilon_1$ with other results also suggested that our estimate of $\sigma_1$ was more reliable. 

Figure 5(b) also suggested that $\sigma_1(T)$ of Na-DNA was strongly decreased with decreasing temperature above about 100~K, as was expected for a semiconducting material. 
As shown in Fig.~6(a), the Arrhenius plots of the normalized conductivity by the value at 300~K suggested that the thermally activated behavior with an energy gap of 62~meV; $\sigma_1\propto e^{(-\Delta/k_BT)}$, was observed at least down to 200~K, independent of the samples. 
However, it seemed that the behavior of $\sigma_1(T)$ below 100~K was saturated and dependent on the samples. 
When we subtracted the value of $\sigma_1$ at $T$=100~K from $\sigma_1(T)$, we found that the thermally activated behavior with the same energy gap could be expanded down to about 100~K, as shown in Fig.~6(b). 
There are several possibilities to be considered as the origin of the saturated behavior of $\sigma_1(T)$ below 100~K, such as the residual contribution of the glass tube, the counterions and the impurities. 
In any case, we can conclude that the charge transport of Na-DNA at least near 300~K is determined by a thermal activation mechanism. 

In general, the following three channels can be considered as the charge transport mechanism of DNA samples near 300~K; (i) an intrinsic contribution of charge along DNA strands, (ii) a relaxational motion of the surrounding water molecules, and (iii) an ionic conduction of counterions. 
First of all, it seems that the contribution of counterions is almost negligible near 300~K, since the temperature dependence of $\sigma_1(T)$ for our Na-DNA near 300~K was sufficiently stronger than that of a lyophilized buffer for DNA samples, which were measured by Tran {\it et al}.\cite{TranPRL00} 
As was discussed by Tran {\it et al.}, the contribution of counterions may become more important below 100~K. 
Next, Briman {\it et al}.\cite{Briman} have recently demonstrated that even if a DNA sample was prepared at almost negligible humidity, the fast relaxation of single water molecule attached to the DNA sample can contribute to the conductivity in the frequency range above 100~GHz. Thus, we compared the contribution of this effect with our results at our measured frequency, as shown in Fig.~7. 
It is clear that our results were at least one order of magnitude larger than the value at 10~GHz of the contribution from single water molecule relaxation, 
strongly suggesting that the influence of the water molecule relaxation is almost negligible in our samples. Therefore, we concluded that the observed semiconducting behavior of Na-DNA near 300~K was very close to the intrinsic property of the charge transport along DNA strands. 

According to recent first-principles numerical simulations based on the dynamic functional theory (DFT), the electronic structure of a stacked nucleic base pairs (at least up to 30 pairs) has an energy gap of the order of 1$\sim$10~eV between HOMO and LUMO, suggesting that DNA is essentially a band-gap insulator \cite{Gervasio,Artacho,Endres}. 
Of course, the real value of the band gap is still unknown, since DFT calculations tend to underestimate the band gap of insulators \cite{Artacho}. 
Moreover, the electronic structure of base pairs is strongly affected by the environment around base pairs, such as phosphate backbone, counterions, and solvating waters \cite{Gervasio,Artacho}. 
However, these numerical results do not seem to support the early suggestion by Eley and Spivey \cite{Eley1962}, who suggested that the $\pi-\pi$ coupling of base pairs in double-stranded DNA could lead to conducting behavior. 
Thus, it seems that another model including a possible mechanism of carrier doping into DNA is required to explain the semiconductive behavior we observed. 

In our samples, DNA strands are considered to be strongly winding by vacuum drying treatment. 
Such drastic deformation of DNA may induce local imbalance in the charge neutrality between negatively-charged phosphates and counter-cations, implying a possibility that electrons or holes can be doped locally. 
In this case, both electrons and holes are doped by the same amount, and their electrical conduction is essentially ambipolar. However, it is more important that the doped carriers can migrate from the outside region around phosphates to DNA base pairs, since the motion of carriers remaining in the outside region may be indistinguishable from the ionic conduction. 
At present, it is unclear whether such migration of carriers occurs or not. 

On the other hand, the large thermal fluctuations of DNA structure and counterion position should be also considered, although they have not been investigated sufficiently by the DFT simulations. 
A classical molecular dynamics simulation suggested that the Na$^+$ counterions in wet DNA populated not only the region near phosphates but also the grooves of DNA helix, because of the thermal fluctuations of counterions \cite{Barnett}. 
The Na$^+$ counterions residing in the groove was considered to inject holes into HOMO band more effectively. 
In our vacuum dry samples, the deformation of DNA strands may trap an amount of counterions in the groove of helix. 
In particlular, the existence of dehydrated Na$^+$ by drying process is expected to play the role of efficient dopants \cite{Kino}. 

Interestingly, the hole doping into HOMO band seems to favor the observed semiconductive behavior with a relatively small activation energy, $E_a$. 
A recent {\it ab initio} study based on DFT calculations found clear evidence for the formation of small polarons in a hole-doped segment of dry four cytosine-guanin pairs \cite{Alexandre2003}. 
The polaron binding energy, $E_b$, was calculated to be $\sim$0.3~eV, nearly independent of the fragment size. 
In a small polaron model, the electrical conduction at high temperatures is dominated by thermal hopping of the localized polarons with $E_a$ which is substantially smaller than $E_b$ \cite{Holstein}. 
Thus, the observed thermally activated behavior with $E_a$ less than 0.1~eV seems to be consistent with the small polaron model due to doped holes into HOMO band. 
A recent study of dc $I$-$V$ characteristics for DNA molecules with several thousands of identical base pairs also reported similar results which could be explained by the polaron hopping model \cite{Yoo2001}. 

Unfortunately, the experimental verification of the above speculation on carrier doping is beyond the scope of the present work. 
We also note that the estimation of $\sigma_1$ in this work is only a three dimensional averages of the quasi-one dimensional conductivity along DNA strands, since our DNA strands were strongly shrinked. 
Nevertheless, our results showed that the observed semiconductive behavior was very close to the intrinsic property of the charge transport of DNA molecules rather than the dipole relaxation of residual water molecules and the ionic conduction of counterions. 
In order to obtain further insights into the charge transfer mechanism in DNA, a more sophisticated preparation of DNA samples and an improvement of the microwave conductivity measurement are crucially required. 

\section{Conclusion}

The systematic microwave measurements were performed to obtain both $\sigma_1(T)$ and $\epsilon_1(T)$ of Na-DNA, by enclosing sample powders in the glass tubes together with dry gas in order to suppress the influence of the residual water molecules. 
Both the magnitude and the temperature dependence of $\sigma_1(T)$ and $\epsilon_1(T)$ strongly suggested that the electrical conduction of Na-DNA is essentially semiconductive, and that this behavior was very close to the intrinsic property of the charge transport along DNA strands. 

\section*{Acknowledgment}
We thank H. Fukuyama for valuable discussions and comments, 
M. Edamatsu and Y. Toyoshima for useful advice about DNA samples. 
This work was partly supported by the Grant-in-Aid for Scientific Research 
(1565405) 
from the Ministry of Education, Science, Sports and Culture of Japan.

\begin{table}[b]
 \caption{Parameters of the measured sample tubes. 
 Here, $M_{\rm DNA}$, $V_{\rm DNA}$, $\delta$, $D_{\rm tube}$, and 
 $H_{\rm tube}$ are a mass of DNA, an apparent volume of DNA packed 
 in the tube, a packing fraction determined by a ratio of a true volume to the apparent volume, an outer diameter of the glass tube, a height of the glass tube, respectively. See the text for details of sealing method.}
 \label{tbl1}
  \begin{tabular}{ccccccc}
  \hline
  sample  & $M_{\rm DNA}$ (mg) & $V_{\rm DNA}$ (mm$^3$) & $\delta$ & $D_{\rm tube}$ (mm) & $H_{\rm tube}$ (mm) & sealing method \\
  \hline
   A & 0.713 & 2.92 & 0.15 & 1.6 & 17.4 & I  \\
   B & 0.305 & 0.77 & 0.27 & 1.6 & 17.1 & I  \\
   C & 1.784 & 2.67 & 0.41 & 1.2 & 12 & II  \\
   D & 1.088 & 1.91 & 0.35 & 1.2 & 12 & II  \\
   E & 0.401 & 0.69 & 0.36 & 1.2 & 12 & II  \\
   empty tube & --- & --- & --- & 1.6 & 14.7 & I \\
  \hline
  \end{tabular}
\end{table}

\begin{figure}[t]
\begin{center}
\includegraphics[width=4in]{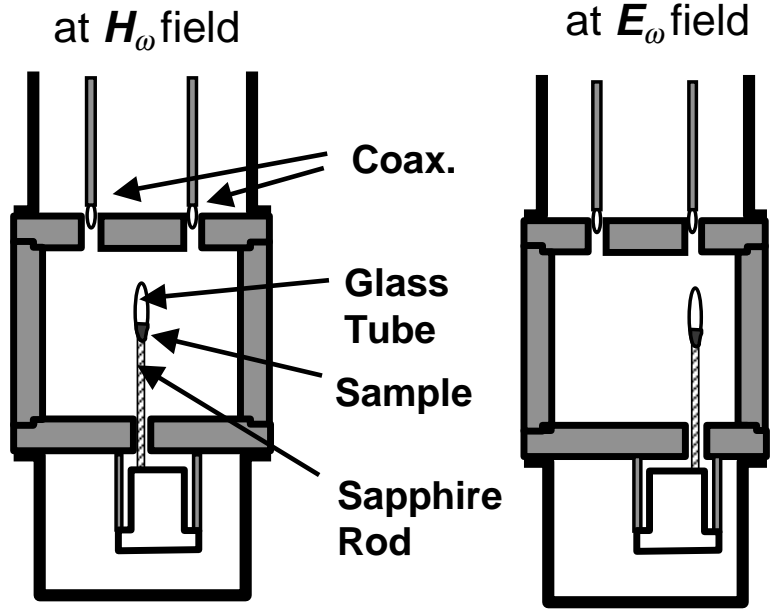}
\end{center}
\caption{A schematic viewgraph of the cylindrical cavity resonator and the sample tube measured. Two sample positions are selected to measure the microwave response at $H_\omega$ or $E_\omega$ field. }
\label{f1}
\end{figure}

\begin{figure}[t]
\begin{center}
\includegraphics[width=12cm]{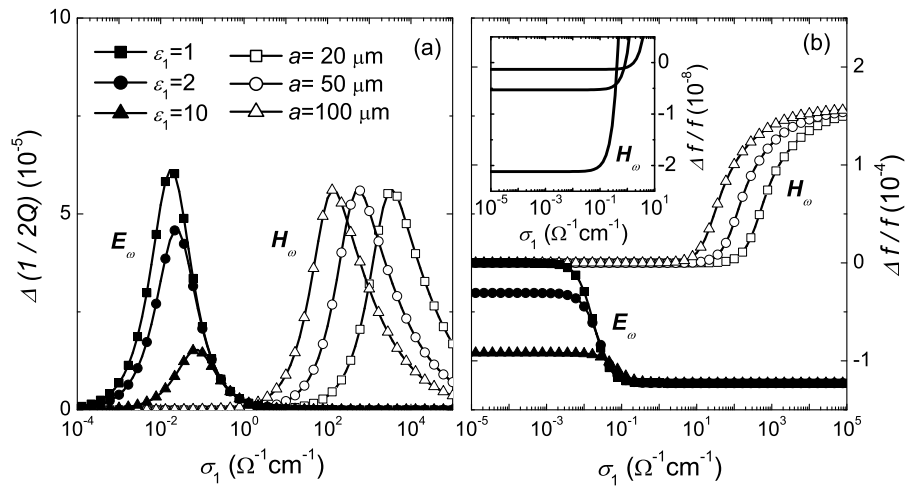}
\end{center}
\caption{(a) Calculated behavior of (a) $\Delta(1/2Q)$ and (b) $\Delta f/f$ at $E_\omega$ (lines with solid symbols) and $H_\omega$ (lines with open symbols) as a function of the real part of the complex conductivity, $\sigma_1$. Inset: An expanded viewgraph of $\Delta f/f$ at $H_\omega$ in the lower conductive region.}
\label{f2}
\end{figure}

\begin{figure}[t]
\begin{center}
\includegraphics[width=12cm]{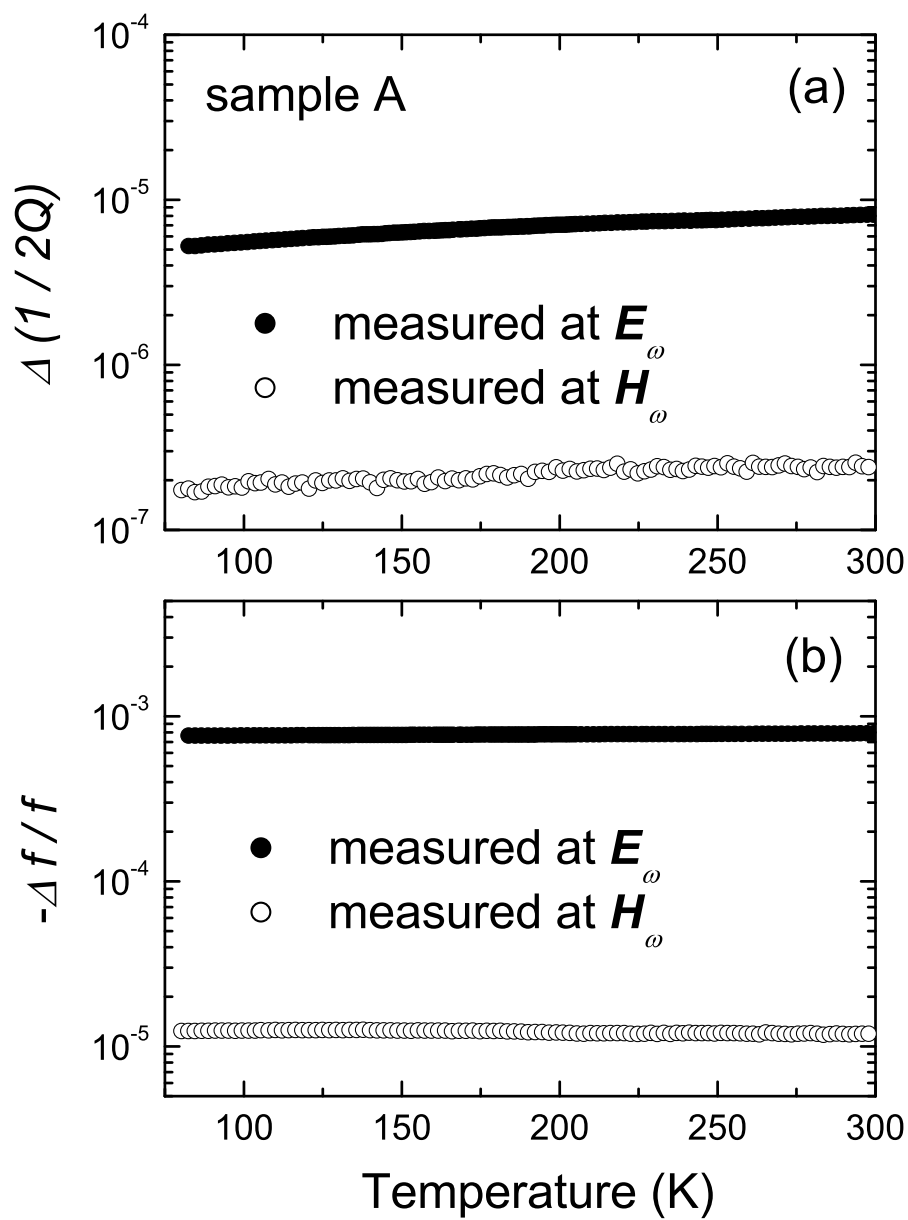}
\end{center}
\caption{(a) The comparison of $\Delta(1/2Q)$ at $E_\omega$ and $H_\omega$ for sample A. (b) The comparison of $-\Delta f/f$ at both field for sample A.}
\label{f3}
\end{figure}

\begin{figure}[t]
\begin{center}
\includegraphics[width=15cm]{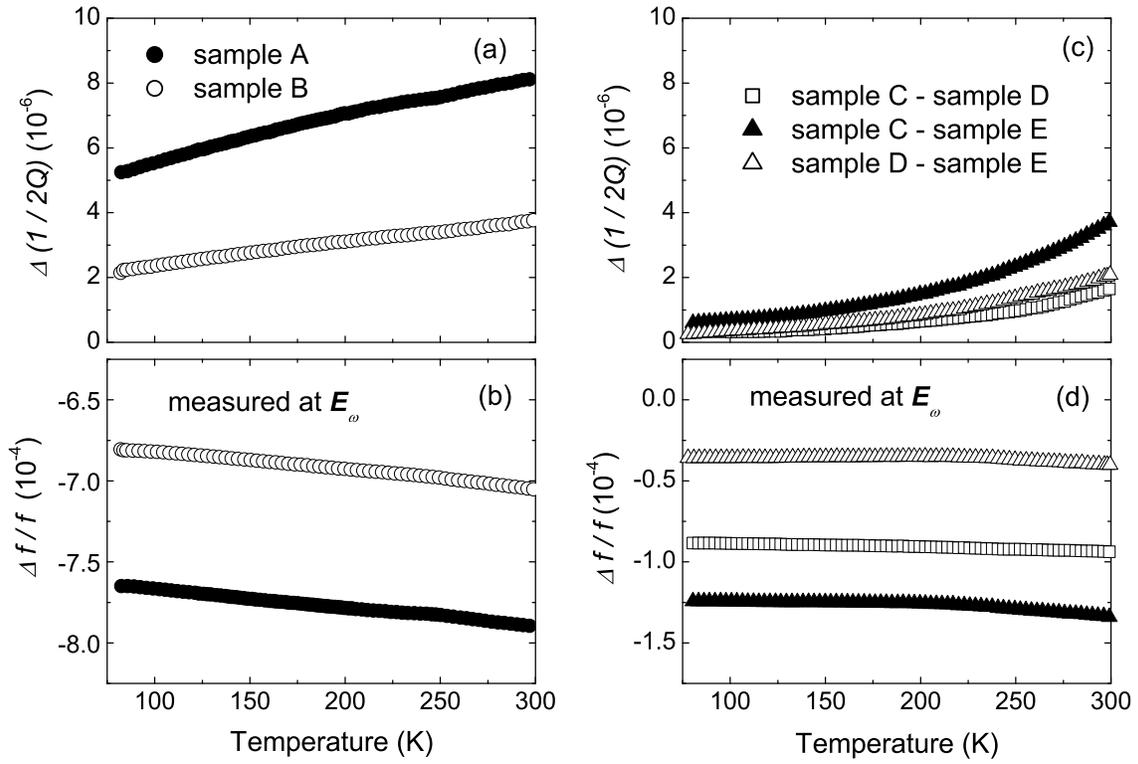}
\end{center}
\caption{(a) $\Delta(1/2Q)$ and (b) $\Delta f/f$ for samples A and B, which were obtained by subtracting the data for the empty tube. (c) $\Delta(1/2Q)$ and (d) $\Delta f/f$ for samples C to E.}
\label{f4}
\end{figure}

\begin{figure}[t]
\begin{center}
\includegraphics[width=10cm]{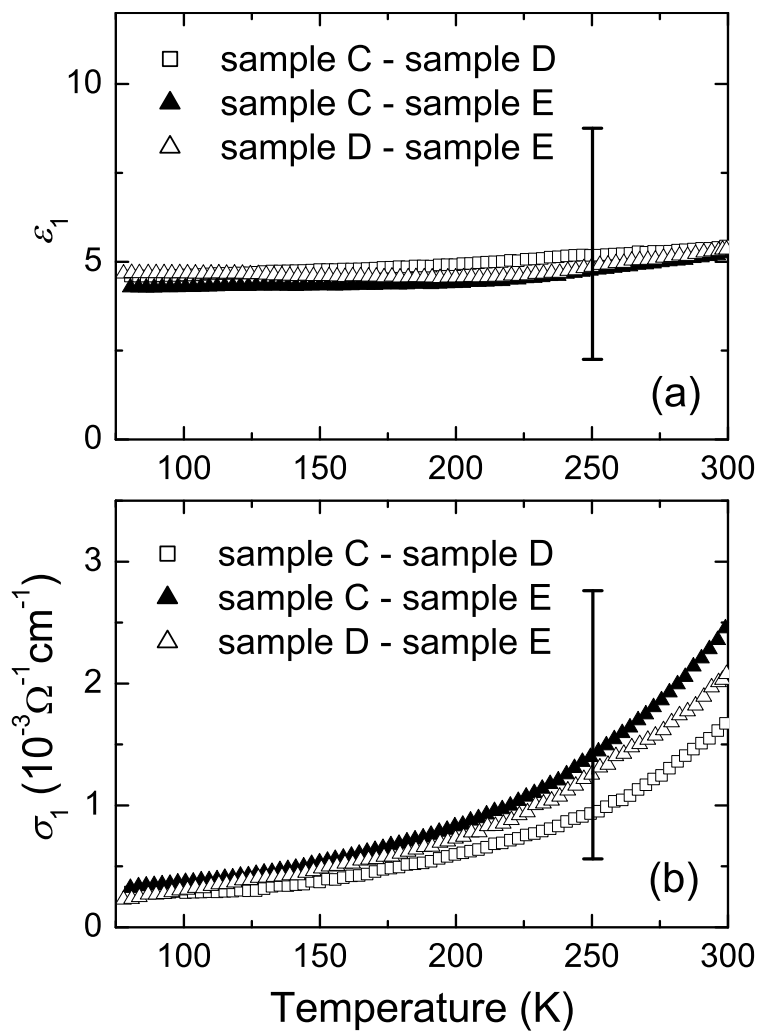}
\end{center}
\caption{(a)The temperature dependence of $\epsilon_1(T)$ for samples C to E. (b) The temperature dependence of $\sigma_1(T)$}
\label{f6}
\end{figure}

\begin{figure}[t]
\begin{center}
\includegraphics[width=10cm]{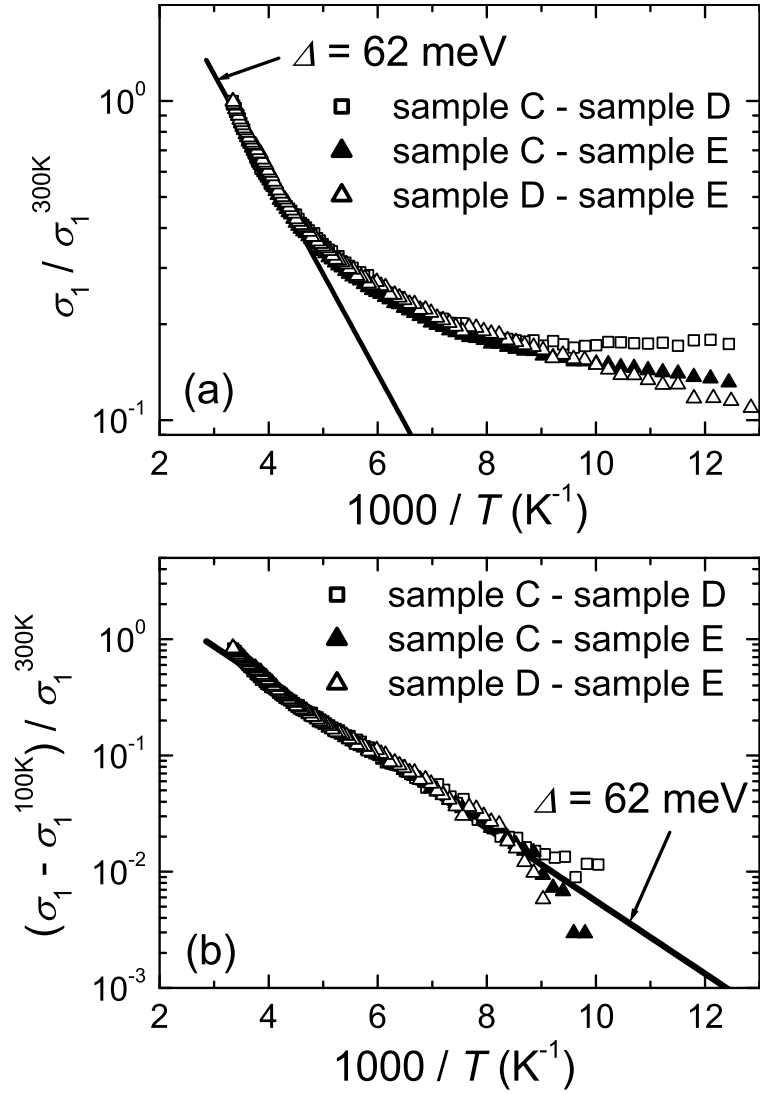}
\end{center}
\caption{(a)The Arrhenius plots of the normalized conductivity by the value at 300~K. (b) A similar plots after subtracting the contribution below 100~K.}
\label{f7}
\end{figure}

\begin{figure}[t]
\begin{center}
\includegraphics[width=12cm]{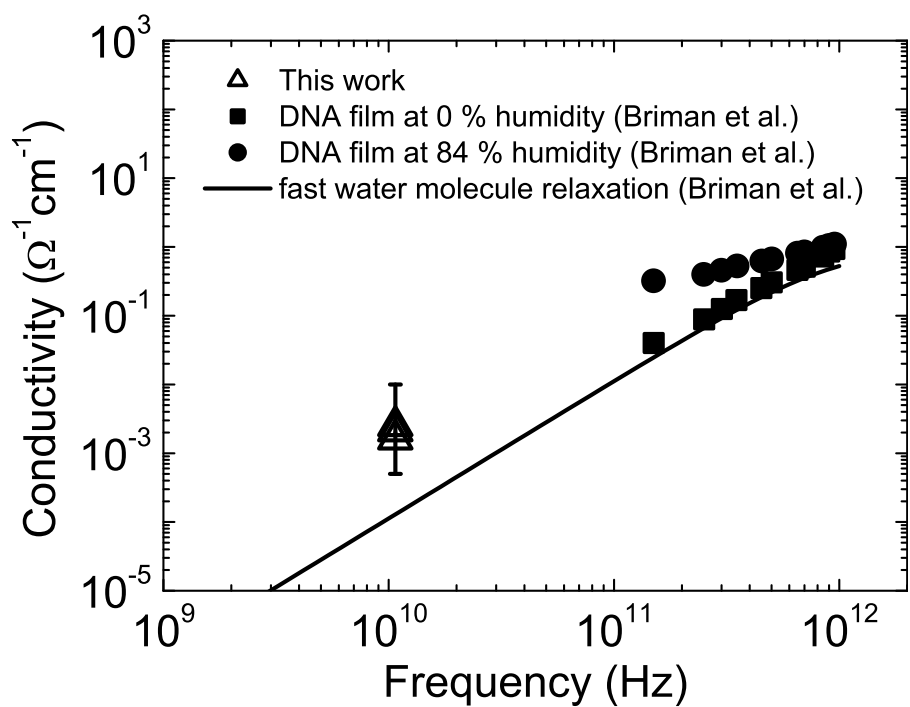}
\end{center}
\caption{Conductivity of DNA as a function of frequency at 300~K.}
\label{f7}
\end{figure}

\end{document}